\documentclass[aps,prl,twocolumn,notitlepage,nofootinbib]{revtex4-1}
\usepackage{makecell}
\usepackage{mathrsfs}
\usepackage{mathbbol}
\usepackage{amsmath}
\usepackage{bm}
\usepackage{graphicx}
\usepackage{color,soul}
\usepackage{float}
\usepackage{multirow}
\usepackage[inline]{enumitem}
\usepackage{cancel}
\usepackage{slashed}
\usepackage[normalem]{ulem}
\usepackage[breaklinks,colorlinks,urlcolor=blue,citecolor=citcolor,linkcolor=lcolor]{hyperref}
\definecolor{lcolor}{rgb}{0.5,0,0}
\definecolor{citcolor}{rgb}{0,0.3,0.0}
\definecolor{ao(english)}{rgb}{0.0, 0.5, 0.0}
\newcommand{\xB}{{x_{\text{B}}} }
\newcommand{\rmd}{{\rm{d}} }

\allowdisplaybreaks

\begin{document}
\title{Modification of the Jet Energy-Energy Correlator in Cold Nuclear Matter}

\author{Yu Fu}
\author{Berndt M\"uller}
\author{Chathuranga Sirimanna}
\affiliation{Department of Physics, Duke University,
Durham, North Carolina 27708, USA}

\begin{abstract}
We compute medium corrections to the energy-energy correlator (EEC) for jets in electron-nucleus collisions at leading order in the QCD coupling and the interaction of the jet with the medium. We derive an analytical expression for the modification of the EEC as a function of the opening angle and show that the modification is strongest at large angles within the jet cone. We obtain explicit results for the dependence of the modification on the jet energy, the scattering power of cold nuclear matter, and the path length within the medium. We extend our calculations to gluon jets in proton-nucleus collisions and compare our results with recent preliminary data for proton-lead collisions at the LHC. We also discuss the role of comovers on the EEC in p+Pb collisions.
\end{abstract}

\maketitle

%%%%%%%%%%
\noindent{\it Introduction.--} Jets are an emergent Quantum Chromodynamics (QCD) phenomenon that can be studied in collider experiments. They encode valuable information about the interactions of high-energy quarks and gluons and their confinement into hadrons. Jets have also emerged as powerful probes of properties of the quark-gluon plasma created in high-energy heavy-ion collisions. Over the past decades, various jet measurements have been conducted in a wide range of collision systems \cite{Rabbertz:2017ssq,Connors:2017ptx,Cunqueiro:2021wls}). Attention has been directed recently at the detailed structure of energy flow within jets, often referred to as jet substructure \cite{Larkoski:2017jix,Kogler:2018hem,Marzani:2019hun}.

A particularly interesting infrared-safe jet substructure observable is the energy-energy correlator (EEC). It measures the total energy deposited in two ideal detectors as a function of the angle between the detectors.  Modern high-energy colliders with the exceptional angular resolution of tracking detectors and calorimeters have enabled the study of EECs in a wide variety of collision systems. These studies hold the promise of improving our understanding of various aspects of these collisions, including imaging of the confining transition from quarks and gluons to free hadrons, probing the scale-dependent structure of the quark-gluon plasma \cite{Andres:2022ovj,Andres:2023xwr,Yang:2023dwc}, exploring the quark mass effect on jet substructure \cite{Craft:2022kdo,Xing:2024yrb}, extracting top quark mass \cite{Holguin:2024tkz}, determining the strong coupling \cite{CMS:2024mlf} and studying nucleon intrinsic dynamics\cite{Liu:2022wop,Cao:2023oef}.

Although EECs have been widely studied for jets in the vacuum and hot nuclear matter, they have been scarcely explored in cold nuclear matter. Such matter can be probed by jets in electron-nucleus (e+A) or proton-nucleus (p+A) collisions. The future electron-ion collider (EIC) will provide unique opportunities to study jet physics in deep-inelastic scattering (DIS) \cite{Accardi:2012qut,Aschenauer:2017jsk,AbdulKhalek:2021gbh}. Various jet observables, such as jet-$p_T$ broadening \cite{Liu:2018trl}, jet angularities \cite{Aschenauer:2019uex}, jet charges \cite{Li:2020rqj}, heavy-flavor tagged jet substructure \cite{Li:2021gjw}, have been investigated with the aim to develop a jet physics program at the EIC. One phenomenological study of EEC in e+A collisions \cite{Devereaux:2023vjz} was based on the eHIJING \cite{Ke:2023xeo} event generator. Recently, the ALICE collaboration at the Large Hadron Collider reported preliminary results on EECs for jets produced in proton-lead collisions at 5.02 TeV.

In this \textit{Letter}, we present the first analytical calculation of the modification of the jet EEC in cold nuclear matter using the higher-twist (HT) formalism at the leading order(LO). We take both initial-state and final-state cold nuclear effects into account and provide the full expression of the energy-weighted differential cross section in DIS as a function of the EEC angle $\theta$. We find that the initial-state effect brings about a suppression of the nuclear EEC ratio in all angular regions, while the final-state interactions between jet and cold nuclear matter lead to an enhancement at large angles within the jet cone.  We further extend our calculation to p+Pb collision and compare our results to preliminary data reported by ALICE.\footnote{The preliminary ALICE data for jet EEC in p+Pb collisions have been presented at a recent conference \cite{Nambrath:2024hp}, but the data set has not yet been publicly released.}
\begin{figure}
     \centering
     \includegraphics[width=0.8\linewidth]{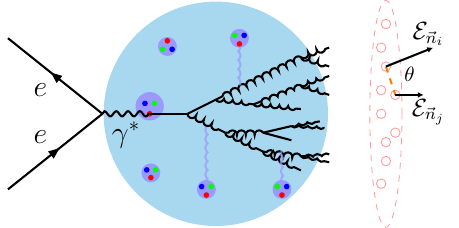}
     \caption{A hard collision, here represented in the Breit frame, between the virtual photon and a quark initiates a parton shower. The EEC measures the energy flow carried by final-state hadrons along the directions $\vec{n}_i$ and $\vec{n}_j$ inside the jet cone.}
     \label{fig::EECinDIS}
 \end{figure}

\noindent{\it  Formalism --}
To explore in-medium modifications of jet substructure in DIS processes, we examine how the energy flow correlation $\langle\mathcal{E}_{\vec{n}_i} \mathcal{E}_{\vec{n}_j} \rangle$ within the jet depends on the angular scale $\cos\theta = \vec{n}_i\cdot\vec{n}_j$ (see Fig.~\ref{fig::EECinDIS}). Jet EECs are understood in two main domains: the small-angle region, where the EEC reflects the distribution of uncorrelated hadrons, and the large-angle region, where it reflects the structure of partonic splittings in perturbative QCD. These domains are separated by a broad peak governed by confinement effects, whose position scales with the total jet momentum $p_T$ as $p_T\theta_{\rm peak} \approx 2.4$ GeV/c \cite{ALICE:2024dfl}. We focus on the perturbatively calculable large-angle region where the cold nuclear medium modifies the partonic splitting pattern inside the jet.

To the LO in the QCD coupling, the energy flow in the jet initiated in a $q\to q+g$ splitting is determined by the momentum and angle distribution of the offspring quark and gluon. We therefore compute the energy-weighted differential cross-section as 
\begin{align}
\frac{\rmd \Sigma}{\rmd \theta} =\int_0^1 \rmd z \frac{\rmd \sigma_{qg} }{\rmd \theta}z(1-z),
\end{align}
where $\sigma_{qg}$ is the inclusive cross-section for a $q\to q+g$ splitting and $z$ is the large momentum fraction carried by the offspring quark. We consider the DIS reaction
\begin{equation}
\text{e}(\ell_1)+ \text{p}/\text{A}(p)  \to \text{e}(\ell_2) + \text{q}(\ell_q) + \text{g}(\ell_g) + \text{X}
\end{equation}
in e+p and e+A collisions, where $\ell_1$ and $\ell_2$ are the four-momenta of the incoming and outgoing electrons, $\ell_q$ and $\ell_g$ are the momenta of the quark and radiated gluon, respectively. $p=[p^+,0^-,\bm{0}_{\perp}]$ is either the momentum of the proton in e+p collision or the momentum per nucleon in the nucleus with the atomic number A in e+A collision. We work in the Breit frame and denote the four-momentum transfer between electron and proton (nucleus) as $q=[-Q^2/2q^-,q^-,\bm{0}_\perp]$.

For a semi-inclusive DIS event, using QCD collinear factorization \cite{Collins:1989gx}, the energy-weighted differential cross-section at LO in $\alpha_s$ for a quark jet in the vacuum can be expressed as
\begin{eqnarray}
\label{eq::xsection_ep}
\frac{\rmd\Sigma_{\tt{ep}}} {\rmd \xB\rmd Q^2 \rmd \theta}
&=&\sum_{q}\frac{e_q^2\alpha^2}{2 Q^2 \xB^2s^2}   L_{\mu\nu}H_0^{\mu\nu}(\xB) f_q( \xB)\mathcal{K}_{\tt{LT}}(\theta)
\nonumber \\
\mathcal{K}_{\tt{LT}}(\theta) &=& \frac{\alpha_s}{2 \pi} C_F \int_0^1 \rmd z \frac{(1-z)z}{\theta} P_{qg}(z) \, .
\end{eqnarray}
Here $s=(\ell_1+p)^2$ is the invariant mass squared of the electron-nucleon system, $\alpha$ and $\alpha_s$ are the electromagnetic and strong coupling constant, respectively, $e_q$ is quark charge, and $\xB= Q^2/2p\cdot q$. $P_{qg}(z)$ denotes the quark-gluon splitting function, and $f_q(x)$ is the collinear quark distribution function. $L_{\mu\nu} $ is the leptonic tensor, and $H_0^{\mu\nu}$ is the Born term for $\gamma^*+q$ scattering. As (\ref{eq::xsection_ep}) shows, the angular dependence of the energy-weighted differential cross-section in e+p collision obeys a $1/\theta$ power law. Resumming higher-order splitting processes modifies the angle dependence to $\rmd \Sigma_{\tt{ep}}/\rmd \theta \sim 1/\theta^{1-\gamma(3)}$, where $\gamma(3)$ is the twist-2, spin-3 QCD anomalous dimension at fixed coupling \cite{Li:2021zcf,Jaarsma:2022kdd}. 

The interaction of the jet partons with the surrounding nucleus modifies the $\theta$-scaling of the EEC in two ways. Firstly, the nuclear environment affects the parton distribution $f_q(x_\textrm{B})$ at the leading-twist level. The LT energy-weighted cross-section $\rmd\Sigma_{\textrm{eA}}^{\textrm{LT}}/\rmd \xB \rmd Q^2 \rmd \theta$ has the same form as (\ref{eq::xsection_ep}) but uses the nuclear PDF (nPDF) $f_q^A(x)$ instead of the nucleon PDF $f_q(x)$.

Secondly, the parton emerging from a hard interaction in an e+A collision undergoes multiple scatterings and medium-induced gluon bremsstrahlung when it passes through the cold nuclear matter. In a large nucleus this power-suppressed HT process is proportional to the number of nucleons encountered by the scattered parton on its way out of the nucleus. This number scales with the nuclear radius $R_A \approx r_0 A^{1/3}$, leading to a nuclear enhancement of the HT contribution \cite{Luo:1994np}. In the HT formalism, the hadronic tensor for a single rescattering with medium-induced gluon emission can be factorized as the convolution of a twist-4 quark-gluon correlator $T_{qg}(x)$ and the second derivative of the hadronic tensor with respect to the rescattering momentum transfer $k_T$  \cite{Guo:2000nz,Wang:2001ifa}. $T_{qg}(x)$ contains both initial and final-state information.

In the collinear limit, the twist-4 correlator $T_{qg}$ can be expressed in terms of the twist-2 nPDF \cite{Casalderrey-Solana:2007xns,Majumder:2010qh} and the jet transport coefficient $\hat{q}$, which is defined as the average transverse momentum broadening per unit length \cite{Baier:1996sk,Guo:2000nz}. The corresponding contribution to the energy-weighted differential cross section at the next-to-leading twist (NLT), $\rmd\Sigma_{\textrm{eA}}^{\textrm{NLT}}/\rmd \xB \rmd Q^2 \rmd \theta$, has the same form as its LT counterpart (\ref{eq::xsection_ep}), but with nPDF instead of PDF, and the replacement
\begin{align}
\label{eq::xsection_NLT}
\mathcal{K}_\textrm{NLT}&(\theta,q^-) = \! \dfrac{\alpha_s}{2\pi }C_A\int_0^1 \rmd z  P_{qg}(z)\frac{1}{\pi } \dfrac{8}{\theta^3z(1-z)(q^-)^2} 
\nonumber\\
&\times \int_0^L \! \!\rmd\xi^- \hat{q}(\xi^-)\, \sin^2 \left( \dfrac{z(1-z)q^-\theta^2}{4}\xi^- \right).
\end{align}
Here $q^- =Q/\sqrt{2}$, $\mathcal{K}_\textrm{NLT}(\theta,Q)$ is the kernel accounting for the parton splitting and in-medium rescattering processes and $L$ is the total path length of the jet inside the nucleus. The most relevant aspect of this result is that the angular dependence of the NLT contribution to the energy-weighted differential cross section is linear in $\theta$ when the argument of the sine function is small and thus behaves inversely to that of the LT contribution (\ref{eq::xsection_ep}).

\noindent{\it Discussion --}  Jet EECs are influenced by cold nuclear matter effects. Depending on whether the medium affects the jet before or after the hard collision, these effects are categorized as either initial-state or final-state effects. In the NLT contribution due to double scattering, the initial-state effect is encoded in the nPDF, which is often expressed as $f_q^A(x,Q^2)=R_q^A(x,Q^2)f_q(x,Q^2)$ with the nuclear modification factor $R_q^A$ \cite{Eskola:2021nhw}. The final-state effect is expressed in terms of $\hat{q}$. Substantial efforts have been made to extract $\hat{q}$ for hot QCD matter \cite{JET:2013cls,Chen:2016vem,Andres:2016iys,Bianchi:2017wpt,Ma:2018swx,Zhou:2019gqk,JETSCAPE:2021ehl,JETSCAPE:2024cqe}.  Efforts aimed at extracting $\hat{q}$ for cold nuclear matter are less advanced because of a relative dearth of useful data. The first global analysis gave an estimate $\hat{q}_{\rm CNM} \approx 0.02$ $\text{GeV}^2/\text{fm}$ \cite{Ru:2019qvz}, which is approximately two orders of magnitude smaller than its value in a hot medium. Here we ignore a possible scale dependence of $\hat{q}$ \cite{Ru:2019qvz, Kumar:2019uvu} and assume that its value over the considered range of $x_B$ and $Q^2$ is constant.

As discussed above, our treatment of single and double scattering in e+A collisions allows us to disentangle the nuclear effects arising from initial-state modifications from those due to final-state interactions. The initial-state effects already occur at LT and introduce an overall nuclear modification factor into the energy-weighted differential cross section. Depending on the kinematic region of interest, this factor can either suppress or enhance the jet EEC in cold nuclear matter compared to those in the vacuum. How this nuclear suppression is reflected in the EEC depends on the normalization of the energy-weighted cross sections. 

On the other hand, as already shown in (\ref{eq::xsection_NLT}), the final-state effect affects the jet EEC in two ways. The scattering power and size of the medium not only affects the magnitude of the nuclear modifications of the EEC but also gives rise to a different angle dependence, which is a combination of the kinematic factor $\theta^{-3}$ and a phase factor $\theta$ that encodes the Landau–Pomeranchuk–Migdal (LPM) effect. As a result of this combination the NLT contribution to the jet EEC increases with $\theta$ within the jet cone. To summarize, the effect of the cold nuclear medium on jet EEC is the combined result of the nuclear modification factor of the PDF, $R_i^A$, and the jet transport coefficient $\hat{q}$. It also depends on the effective path length $L$ of the jet inside the nucleus. 

To better quantify the medium modification of jet EEC in cold nuclear matter, accurate measurements of both $R_i^{A}$ and $\hat{q}$ are required. So far, the extraction of $\hat{q}$ is rather limited  in precision \cite{Ru:2019qvz}. The measurement of $R_i^A$ has been continuously improved over the past decade \cite{Eskola:2008ca,Eskola:2009uj,Eskola:2016oht,Eskola:2021nhw,Schienbein:2009kk,Kovarik:2015cma} and will be further improved by measurements at the future EIC \cite{Aschenauer:2017jsk}. Once this goal is achieved, the measurement of jet EEC in e+A collisions at the EIC will offer a promising avenue to further constrain $\hat{q}$ for cold nuclear matter.

\noindent{\it Results --}
To examine the in-medium modification of jet EEC, it is convenient to consider the ratio of the energy-weighted cross-section in e+A collisions to that in e+p collisions,
\begin{align}
\label{eq::EECratio}
R_\textrm{eA/ep}(\theta)= \frac{\rmd\Sigma_{\textrm{eA}}^{\textrm{LT+NLT}} / \rmd \xB \rmd Q^2 \rmd \theta }{ \rmd\Sigma_{\tt{ep}} /\rmd \xB\rmd Q^2 \rmd \theta} ,
\end{align}
henceforth called the EEC ratio, as a function of the angle $\theta$ at fixed $\xB$ and $Q^2$.
\begin{figure}
    \centering
    \includegraphics[width=\linewidth]{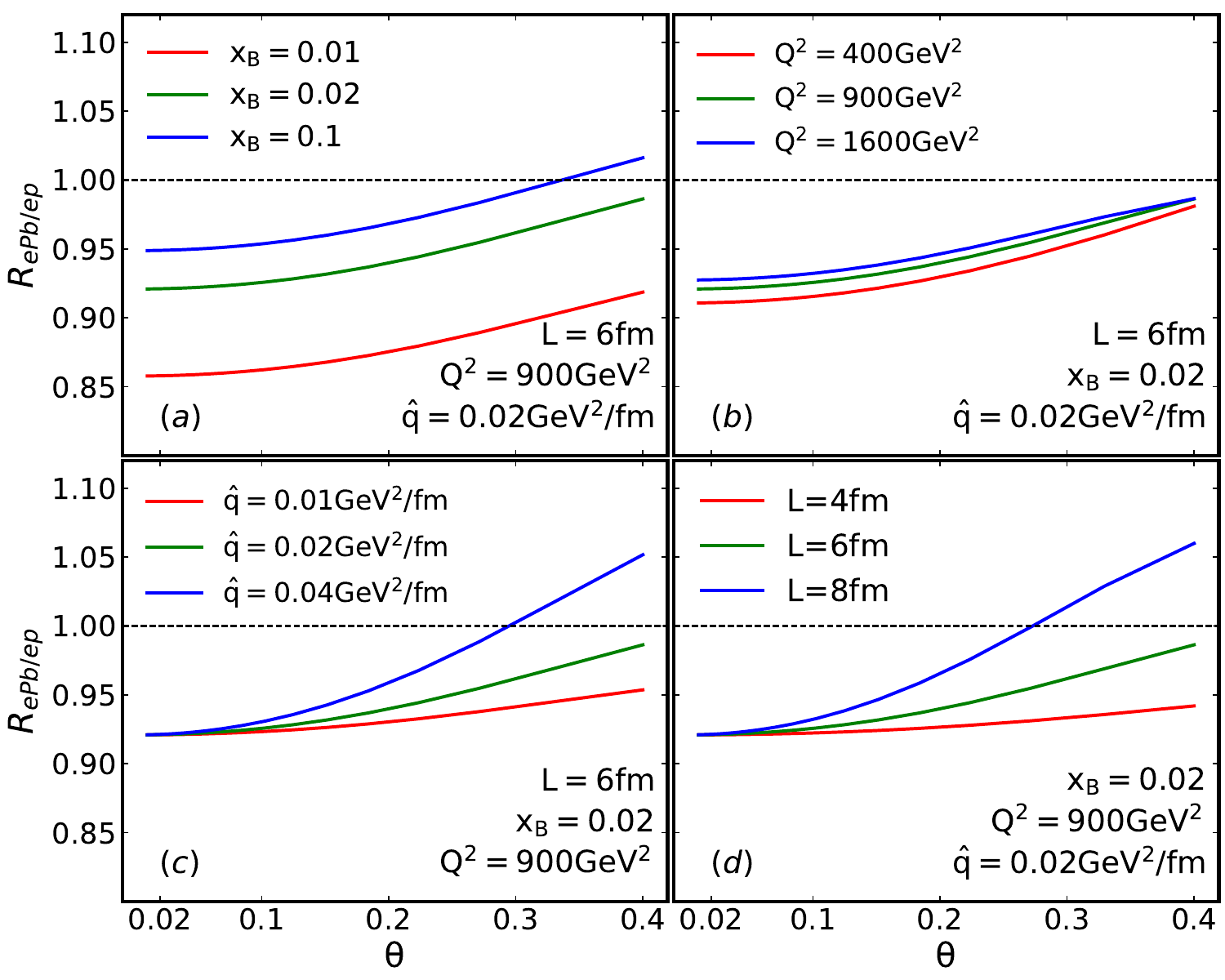}
    \caption{EEC ratio in e+Pb as a function of the angle $\theta$ for various combinations of the four parameters, varying $\xB$, $Q^2$, $\hat{q}$ and $L$ in panels (a-d), respectively.}
    \label{fig::Ratio_eA}
\end{figure}
In our evaluations of (\ref{eq::EECratio}) we use the CT10 fits \cite{Lai:2010vv} for the proton PDF and the nCTEQ15FullNuc fits \cite{Kovarik:2015cma} for the nPDF. The EEC ratio depends on four parameters, $\xB, Q^2, \hat{q}$, and the path length $L$, where the first two parameters are related to the primary hard scattering while the last two encode the final-state interaction of the scattered quark with the nucleus. 

Shown in Fig.~\ref{fig::Ratio_eA} are our numerical results for the EEC ratio in e+Pb for various combinations of the four parameters. As discussed above, the initial-state effect results in an overall, angle-independent modification, while the final-state effects generally lead to an increase in the EEC ratio in the angular region $Q\theta > 2.4$ GeV. We note that for the virtuality $Q = 30$ GeV considered here the EEC is dominated by nonperturbative QCD physics in the region $\theta \lesssim 0.1$.

In Fig.~\ref{fig::Ratio_eA}(a), we fix $\hat{q}=0.02~\text{GeV}^2/\text{fm}$, $L=6~\text{fm}$, $Q=30~\text{GeV}$ and vary the values of $\xB$. The panel shows that the $\theta$-dependence of the nuclear modification is independent of the value of $x_B$, only the constant baseline changes. In the Fig.~\ref{fig::Ratio_eA}(b), we vary the values of $Q^2$ while keeping the remaining parameters fixed. Here we observe that, unsurprisingly for a HT contribution, the final-state effect on the EEC ratio is more pronounced at lower $Q^2$. In Figs.~\ref{fig::Ratio_eA}(c,d), we study the sensitivity of the final-state nuclear effect on $\hat{q}$ and $L$. As expected, increasing values of either $\hat{q}$ or $L$ strongly enhance the final-state effect at fixed $Q$ and $x_B$, resulting in an increasingly strong growth of the EEC ratio for $\theta > 0.1$.

We emphasize again that the EEC is controlled by nonperturbative physics related to quark confinement and hadronization in the kinematic region $Q\theta < 2.4~ \text{GeV}$. Our finding that the EEC ratio in this region is dominated by angle-independent initial-state effects, which is based on a perturbative calculation, likely remains valid even if the EEC itself is nonperturbative, as a $p_T \sim 30~\textrm{GeV}/c$ quark hadronizes after it leaves the nucleus.

The factorization of the nuclear modified cross section (\ref{eq::xsection_NLT}) into a medium-independent hard part $H_0^{\mu\nu}$ and modified initial-and final-state factors $f_q^A$ and $\mathcal{K}_\textrm{NLT}$ suggests that the EEC ratio remains unchanged when $H_0^{\mu\nu}$ is evaluated at NLO in $\alpha_s$. However, the angle-dependent final-state factor $\mathcal{K}_\textrm{NLT}$ could be modified when the two-loop splitting function is used \cite{Mertig:1995ny,Badger:2004uk}.

We now address how measurements of the nuclear modified EEC would complement existing studies of jet modification in cold nuclear matter. The nuclear modification factor for leading hadrons was measured by the HERMES experiment \cite{HERMES:2000ytc} in the virtuality range $1~{\rm GeV}^2 < Q^2 < 10~{\rm GeV}^2$ and has been analyzed in the HT formalism to extract the value of $\hat{q}$ \cite{Wang:2002ri,Deng:2010xv,Chang:2014fba}. Analyses of the transverse momentum broadening of jets in nuclei have also been used to extract $\hat{q}$ values \cite{Ru:2019qvz} from a variety of data in the range $Q^2 \lesssim 200~{\rm GeV}^2$. None of these analyses uses information about the jet substructure. Jet EEC modifications in e+A scattering at $Q^2 \sim 900 ~{\rm GeV}^2$ from the EIC hold the promise of providing additional constraints on $\hat{q}$ and may help assess the validity of the HT framework for nuclear modifications of jets.

\noindent{\it EEC in p+A collisions --} 
Although e+A data would provide for the ``cleanest'' environment to study modifications of the jet EEC in cold nuclear matter, we can apply our theoretical framework to other collision systems that can probe the same physics, such as proton-nucleus (p+A) collisions. However, unlike the electron, the proton is a composite particle. When one of its partons initiates a hard interaction with the nucleus, the remainder of the proton will usually also interact with the nucleus and produce additional particles that populate then final state. Some of these particles can be kinematically near the jet and interact with it. These particles are known as ``comovers'' \cite{Ferreiro:2014bia}. We will first consider the cold nuclear matter modifiations and then turn to possible comover effects on the jet EEC. 

When expressing the energy weighted jet cross-section for p+p or p+A collisions, the PDF of the parton inside the scattering proton must be taken into account. The parton from the proton can also scatter with the nuclear medium before the hard collision, but such interactions will not affect the substructure of the jet. We therefore only need to consider the effect of nuclear rescattering in the jet showering process.

Given the dominance of gluon jets at midrapidity and moderate jet $p_T$ in both p+p and p+A collisions at high-energy hadron colliders, our focus here is on gluon jets. In this case, the kernel $\mathcal{K}_\textrm{NLT}(\theta,p_T) $ accounting for the parton splitting and medium rescattering processes, is given by an expression analogous to (\ref{eq::xsection_NLT}), but with the gluon-gluon splitting function $P_{gg}(z)$ instead of the quark-gluon splitting function $P_{qg}(z)$. 

Before we present our numerical results, we discuss the contribution from comovers created by soft proton-nucleon interactions. In the laboratory frame, the jet travels in the transverse direction at the speed of light. Its average path length within the comover region is determined by the transverse extent of that region, which approximately equals the proton radius. We thus set $L_\textrm{co} \approx 1~{\rm fm}$. 
For an estimate of $\hat{q}_{\rm co}$ we use the value of the comover multiplicity given by Ferreiro \cite{Ferreiro:2014bia} and scale the value of $\hat{q}$ deduced from jet quenching data in central Pb+Pb collisions \cite{JETSCAPE:2024cqe} by the charged multiplicity ratio of p+Pb to Pb+Pb collisions. 
%As detailed in the supplemental material, 
This leads us to the estimate $\hat{q}_{\rm co} \approx 0.9~{\rm GeV}^2/{\rm fm}$, which is consistent with the absence of noticeable jet quenching in p+Pb collisions at the LHC \cite{JETSCAPE:2024dgu}.

\begin{figure}[hbt]
    \centering
    \includegraphics[width=\linewidth]{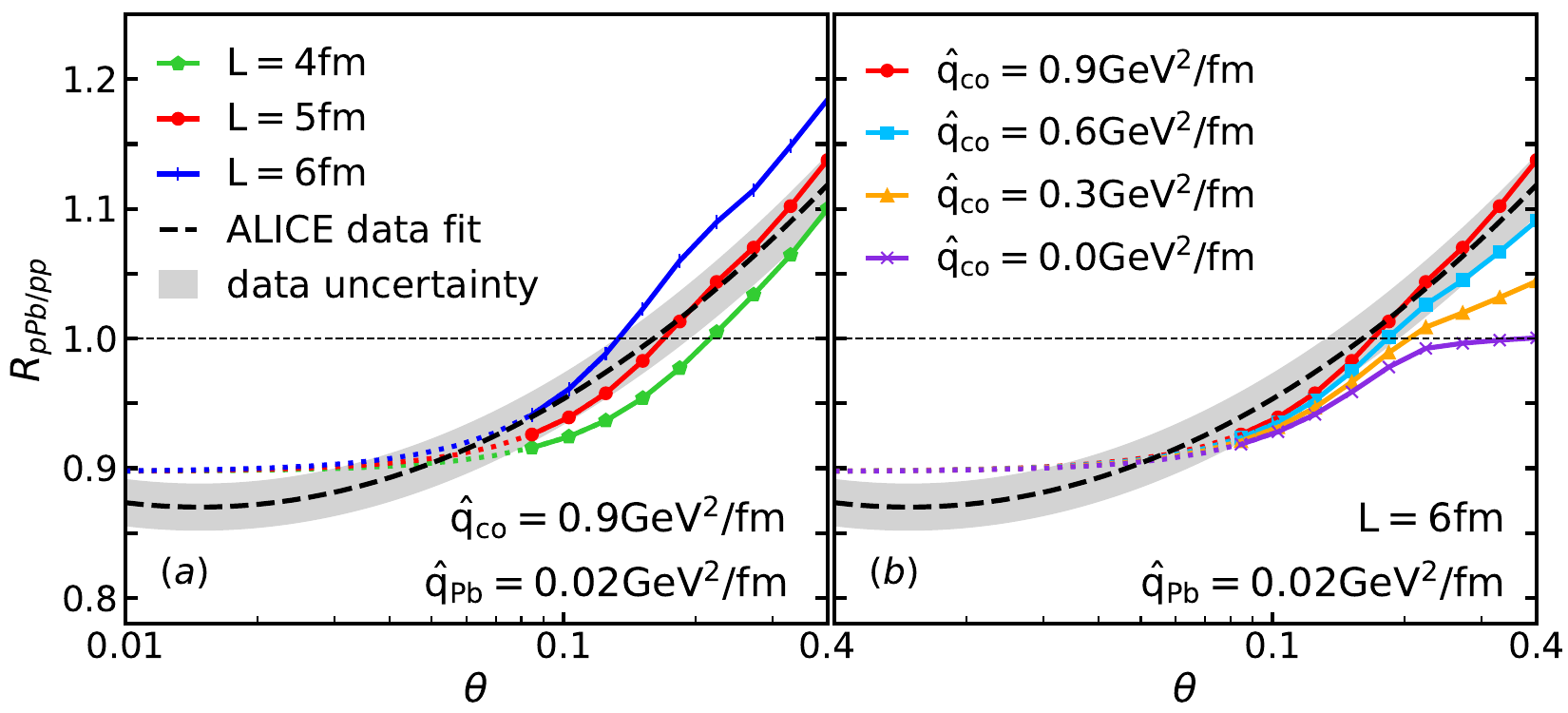}
    \caption{EEC ratio for the 30 GeV jet at midrapidity in p+p/Pb collisions is compared with a fit curve from ALICE data at $\sqrt{s}=5.02$ TeV (dashed black line -- the shaded band around it does not represent the actual experimental uncertainty). The left panel (a) displays the EEC ratio for varying path length $L$; the right panel (b) shows the variation of the EEC ratio for different values of $\hat{q}_{\rm co}$. The solid section of these curves indicates the region of applicability of the perturbative calculation.}
    \label{fig::Ratio_pA}
\end{figure}
We now present our estimates for the nuclear modification of gluon jets in p+A collisions. We again focus on the ratio between the energy-weighted cross-section in p+A collisions and that in p+p collisions:
\begin{align}
R_{\text{pA}/\text{pp}}(\theta)=\frac{\rmd \Sigma_{\tt pA}^{\tt LT + NLT} / \rmd y \rmd p_T^2 \rmd \theta}{\rmd \Sigma_{\tt pp} / \rmd y \rmd p_T^2 \rmd \theta}. 
\end{align}
For a gluon jet at midrapidity with $p_T =30$ GeV/c in p+p (p+Pb) collisions with $\sqrt{s}=5$ TeV, this ratio is plotted as a function of the angle $\theta$ in Fig.~\ref{fig::Ratio_pA}. For the comover effect occurring in a region with a length of 1 fm, we use the estimate for $\hat{q}_{\rm co}$ derived above. Outside this region, $\hat{q}$ is reduced to the cold nuclear matter value $0.02~{\rm GeV}^2/{\rm fm}$ for a total length $L$. A fitted curve based on preliminary ALICE data is shown for comparison by adjusting values of the path length $L$. The left panel of Fig.~\ref{fig::Ratio_pA} shows the dependence of the EEC ratio on the pathlength $L$ for a fixed value of $\hat{q}$; the right panel shows the dependence on $\hat{q}_{\rm co}$ for fixed $L$. We emphasize once more that our perturbative calculation is only valid in the region $\theta \gtrsim 0.08$, corresponding to the solid curves in Fig.~\ref{fig::Ratio_pA}. In this region the nuclear shadowing effect in the nPDF causes an overall suppression of the EEC ratio, which is opposed by an enhancement caused by final-state rescattering in the nucleus that grows with increasing $\theta$. At angles approaching the jet cone radius the enhancement from final-state interactions overtakes the suppression due to initial-state shadowing.

If we hypothetically extend our calculation to $\theta \approx 0.01$, a suppression factor of $R_\textrm{pA/pp} \approx 0.89$ arises solely from the nPDF, potentially explaining the EEC ratio at small angles. We tentatively conclude that the suppression of the ratio $R_\textrm{pA/pp}$ at small angles can be attributed to nuclear shadowing.

% \noindent{\it Conclusion and outlook --}
\noindent{\it Summary --} 
We presented the result of the first complete calculation of nuclear modifications of the jet EEC in electron-nucleus collisions at the NLT level. We identified the effects related to initial-state interactions and those arising from final-state interactions between the showering jet and the cold nuclear medium. The initial-state effects are represented by nPDFs, while final-state effects arise from in-jet transverse momentum broadening and the LPM effect. The medium modifications are parametrized by the nuclear modification factor of the PDF, $R_i^A$, and the jet transport coefficient $\hat{q}$ combined with the effective path length $L$ of the jet inside the nucleus.  $R_i^A$ leads to a global modification of the EEC across all relevant angular regions, while the final-state rescattering brings about an enhancement of EEC at angles $\theta \ge 0.1$.  

We also applied the HT formalism to jets in p+A collisions and compared the EEC ratio to preliminary data from the ALICE collaboration. Our results are consistent with the measured angle dependence of the EEC ratio. The initial-state nuclear shadowing effect results in a uniform suppression of the EEC ratio across all angular regions, while final-state rescattering effects become pronounced at larger angles. Notably, at relatively large angles, the enhancement due to final-state interactions surpasses the suppression caused by initial-state shadowing. The inclusion of comovers results in better agreement with the preliminary data. We note that the EEC ratio is more sensitive to comovers than the nuclear suppression factor $R_\text{pPb}(p_T)$ for inclusive hadrons.

Our results represent a step towards advancing the jet physics program at EIC. Measuring the angular behavior of the EEC ratio can provide valuable insights into the extraction of $R_i^A$ and $\hat{q}$, as the EEC is more sensitive to the former at small angles and more sensitive to the latter at large angles within the jet cone. We conclude that the nuclear EEC ratio can serve as a valuable observable in the EIC physics program.

\noindent{\it Acknowledgment.-- } We are grateful to Wenqing Fan, Barbara Jacak, Abhijit Majumder, Ian Moult, Anjali Nambrath, and Ismail Soudi for helpful discussions. This work was supported by the grant DE-FG02-05ER41367 from the U.S. Department of Energy, Office of Science, Nuclear Physics. C.S. is supported in part by the National Science Foundation under grant numbers ACI-1550225 and OAC-2004571 (CSSI:X-SCAPE) within the framework of the JETSCAPE collaboration.

%%%%%%%%%%
%%%%%%%%%%

\bibliographystyle{h-physrev5} 
\bibliography{refs}

\end{document}